\documentclass[prd,twocolumn,showpacs,showkeys,preprintnumbers,floatfix,superscriptaddress,longbibliography]{revtex4-1}

\usepackage{amsfonts} 
\usepackage{amssymb} 
\usepackage{amsmath} 
\usepackage{graphicx} 
\usepackage{subfigure} 
\usepackage{array} 
\usepackage{dcolumn} 
\usepackage{bm} 
\usepackage{latexsym} 
\usepackage{longtable} 
\usepackage{hyperref} 
\usepackage{bbold}
\usepackage{mathtools}
\usepackage{algorithm}
\usepackage[noend]{algpseudocode}

\usepackage[utf8]{inputenc}
\usepackage[braket]{qcircuit}
\usepackage[dvipsnames,usenames]{xcolor}

\begin{document}
\title{Entanglement and Many-Body effects in Collective Neutrino Oscillations}

\author{Alessandro Roggero}
\affiliation{InQubator for Quantum Simulation (IQuS), Department of Physics,University of Washington, Seattle, WA 98195, USA}

\date{\today}

\begin{abstract}
Collective neutrino oscillations play a crucial role in transporting lepton flavor in astrophysical settings, such as supernovae, where the neutrino density is large. In this regime, neutrino-neutrino interactions are important and simulations in the mean-field approximation show evidence for collective oscillations occurring at time scales much larger than those associated with vacuum oscillations. 
In this work, we study the out-of-equilibrium dynamics of a corresponding spin model
using Matrix Product States and show how collective bipolar oscillations can be triggered by many-body correlations if appropriate initial conditions are present. 
We find entanglement entropies scaling at most logarithmically in the system size suggesting that classical tensor network methods could be efficient in describing collective neutrino dynamics more generally. These observation provide a clear path forward, not only to increase the accuracy of current simulations, but also to elucidate the mechanism behind collective flavor oscillations without resorting to the mean field approximation.
\end{abstract}

\maketitle
 
Neutrinos play a pivotal role in extreme astrophysical events like core-collapse supernovae and neutron star mergers, where they are responsible for both reinvigorating a stalled shock-wave and controlling the conditions for nucleosynthesis in the ejected material~\cite{Hoffman1997,Janka2012,Winteler_2012,Wanajo_2014}. In these environments with a large neutrino density $\rho_\nu$, neutrino flavor evolution is substantially modified by neutrino-neutrino scattering processes which can lead to self-sustained collective flavor oscillations~\cite{Savage1990,Pantaleone92,PANTALEONE1992,Pastor2002B,Balantekin_2005,Fuller2006,Duan2006,Friedland2010,MARTIN2020,Wu2017}. Since neutrinos in supernovae are emitted with fluxes and spectra that are strongly flavor dependent~\cite{Janka2012}, the presence of collective flavor oscillations could then lead to important effects~\cite{Fogli2007,Qian1993,Qian1995}. In some situations however (see eg.~\cite{Duan_2011,Dasgupta_2012}) these could appear only at too large radii to modify the explosion mechanism in supernovae.

For large neutrino densities near the proto-neutronstar, flavor oscillations can occur at frequencies of the order of the neutrino-neutrino interaction energy $\mu=\sqrt{2}G_F\rho_\nu$ instead of the much smaller vacuum oscillation frequency $\omega = \Delta m^2/(2E)$, a phenomenon known as fast flavor conversion~\cite{Sawyer2005,Sawyer2016,Chakraborty2016,Izaguirre2017} (see~\cite{Chakraborty2016b} for a review). In these expressions $G_F$ is Fermi's constant, $E$  the neutrino energy and $\Delta m^2=m_2^2-m_1^2$ the squared mass gap in the two flavor approximation used here.

The standard approach to study these collective phenomena is to consider only the leading order forward-scattering neutrino-neutrino interactions where only flavor can be exchanged among neutrinos (for the full description see eg.~\cite{Vlasenko2014,CIRIGLIANO_2015}). In this scheme, a system with $N$ neutrinos is represented as a collection of $SU(2)$ flavor isospins governed by the Hamiltonian~\cite{Pehlivan2011}~\footnote{Note the difference in the definition of $\mu$ there.}
\begin{equation}
\label{eq:fs_hamilt}
H_{fs} = \sum_{k=1}^N \frac{\Delta m^2}{4E_k} \vec{B} \cdot \vec{\sigma}_k + \frac{\mu}{2N}\sum_{i<j}^N \mathcal{J}_{ij} \vec{\sigma}_i\cdot\vec{\sigma}_j\;,
\end{equation}
with $\vec{\sigma}_i = (\sigma^x_i,\sigma^y_i,\sigma^z_i)$ the vector of Pauli matrices acting on the $i$-th neutrino.
In the expression above, $E_k$ are the neutrino energies and, in the flavor basis $\{\ket{\uparrow},\ket{\downarrow}\}=\{\ket{\nu_e},\ket{\nu_x}\}$, the orientation of the (normalized) vector $\vec{B}=(\sin(2\theta),0,-\cos(2\theta))$ is related to the mixing angle $\theta$. The geometry of the problem is encoded in the two-body coupling matrix $\mathcal{J}_{ij}=\left(1-\hat{p}_i\cdot\hat{p}_k\right)$ with $\hat{p}_k=\vec{p}_k/|\vec{p}_k|$ and $\vec{p}_k$ the momentum of the $k$-th neutrino. Finally, note that this Hamiltonian is obtained under the assumptions of infinite interaction times and that neutrinos have infinite plane-wave solutions, conditions that might not always be appropriate (see eg.~\cite{Stirner_2018}). 
This Hamiltonian model is the base of all the past studies of (two-flavor) collective neutrino oscillations in anisotropic but homogeneous settings (see~\cite{Pehlivan2011} for an explicit derivation of the traditional evolution equations in the mean field approximations from Eq.~\eqref{eq:fs_hamilt})  . The inclusion of spatial inhomogeneity in the neutrino density distribution, important to describe more realistically the environment  encountered in a supernova explosion, will be the subject of future work. The focus here is improving the understanding of the simpler homogeneous case as a fundamental stepping stone for more realistic simulations of flavor dynamics in astrophysical environments.  

Collective neutrino oscillations are understood as being caused by unstable modes in the mean field dynamics generated by the Hamiltonian in Eq.~\eqref{eq:fs_hamilt} which can amplify initially small flavor perturbations exponentially fast~\cite{sawyer2004classical,Duan2010,Izaguirre2017}. The role of entanglement and quantum effects in the out-of-equilibrium dynamics~\cite{Eisert_2015} of neutrinos has received renewed interest recently~\cite{Cervia2019,Rrapaj2020} but is still not well understood, with seemingly conflicting results in the past predicting either a vanishingly small contribution in the large system size limit~\cite{Friedland2003,Friedland2006} or substantial flavor evolution over time scales $\tau_F=\mu^{-1}\log(N)$, which can remain relevant for large systems~\cite{Bell2003,sawyer2004classical}. We will refer to oscillations on the time scale $\tau_F$ to be ``fast'' in contrast to ``slow'' oscillations occurring for $\tau_L=\mu^{-1}\sqrt{N}$. This convention is different from the more common definition in the literature where ``fast'' and ``slow'' oscillations are associated with time scales $\approx\mu^{-1}$ and $\approx\sqrt{\mu\omega}$ respectively, for any system size.

In this work, we study the model Eq.~\eqref{eq:fs_hamilt} for large systems with up to $N=128$ neutrino amplitudes in a simplifying limit and show how coherent flavor oscillations at the fast scale $\tau_F$ can occur even without vacuum mixing. These first-principle simulations, obtained without resorting to the mean-field approximation or exploiting the symmetries present in the model, are made possible by using a Matrix Product State (MPS)~\cite{Vidal2003_mps} representation of the many-body neutrino state. This technique (described in detail in Sec.~\ref{sec:MPS} below) allows for efficient simulation of the real-time dynamics in situations where the entanglement entropy (which provides a measure of quantum correlations)  is relatively small and growing only as the logarithm of the system size. In the context of collective neutrino oscillations, this is potentially very beneficial for two main reasons: first, MPS calculations are a promising route to perform full many-body simulations of neutrino dynamics in medium to large system sizes in order to understand the impact of commonly used simplifications like the mean-field approximation, second, and possibly more importantly, the direct connection between the simulation efficiency of MPS calculations and the amount of quantum correlations generated by the neutrino dynamics can be used as a direct probe of the role of entanglement and its time evolution in the instabilities that lead to the presence of collective modes in neutrino oscillations. An example of the possible insights that can be gained from the entanglement properties of a neutrino system is the connection between dynamical phase transitions and the presence of collective oscillations discussed in Ref.~\cite{Roggero2021b}. Those result were made possible by using  the MPS techniques introduced in the present work and have the potential to complement criteria like linear stability analysis~\cite{Izaguirre2017} in the search of conditions for collective modes to be active in complex astrophysical scenarios like supernova explosions.

The rest of this paper is organized as follows. In Sec.~\ref{sec:model} we introduce the details of the model and the initial conditions used in this work. Sec.~\ref{sec:MPS} provides an introduction to MPS simulations and the techniques used to approximate the time evolution under the neutrino Hamiltonian in Eq.~\eqref{eq:fs_hamilt}. The results of the simulations and described in Sec.~\ref{sec:res} and in Sec.~\ref{sec:conc} we provide a summary and future perspectives. The Appendices contain a direct comparison with the mean field approach and a discussion of the numerical accuracy of our implementation.

\section{Neutrino Spin Model}
\label{sec:model}

We consider a simple model, in the mass basis, in which the neutrinos are divided into two equally sized "beams" ($A$ and $B$) and initialized as the product state $\ket{\Psi_0} = \left(\bigotimes_{n=1}^{N/2} \ket{\downarrow} \right)\otimes\left(\bigotimes_{m=1}^{N/2} \ket{\uparrow} \right)$.
This corresponds to choosing all the neutrinos in the $A$ beam to be $\ket{\nu_2}$ and all the neutrinos in the $B$ beam to be $\ket{\nu_1}$. We will consider the simple case where the energies of the neutrinos in beam $A$ are all equal to $E_A$ and similarly for the neutrinos in beam $B$. Finally, we will neglect the geometric information encoded in the coupling matrix $\mathcal{J}_{ij}$ and take the single angle approximation by setting $\mathcal{J}_{ij}=1$ for all neutrinos~\cite{Duan2010,Pehlivan2011} (see also~\footnote{Since $[H,J^2_A]=[H,J^2_B]=0$ we are free to change the intra-beam couplings at will thus changing the geometry. In the general case we can substitute $\mu\to\mu(1-\cos(\theta_{AB}))$, with $\theta_{AB}$ the angle between the momenta in the $A$ and $B$ beams, for all the results in this work.}). Note that these are not necessarily realistic conditions, as neutrinos are produced in flavor states, but nevertheless contain many of the features relevant for the astrophysical problem. As we will see, the choice of using the mass basis is also useful to better expose differences with the mean-field treatment.

The resulting neutrino Hamiltonian, similar to the one considered in Ref.~\cite{Cervia2019}, can be written as follows
\begin{equation}
\begin{split}
\label{eq:ham_massb}
H &= -\frac{\omega_A}{2}\sum_{i\in\mathcal{A}}\sigma_i^z-\frac{\omega_B}{2}\sum_{i\in\mathcal{B}}\sigma_i^z + \frac{\mu}{2N}\sum_{i<j}\vec{\sigma}_i\cdot\vec{\sigma}_j\;,
\end{split}
\end{equation}
where $\mathcal{A}$ and $\mathcal{B}$ are sets of indices for the neutrinos in the $A$ and $B$ beam respectively while we set the one body energies as $\omega_k=\Delta m^2/(2E_k)$. Since our initial state is completely polarized along the $z$-axis, the mean field approximation for the dynamics generated by $H$ predicts no flavor evolution at all: any time evolution of the flavor polarization is completely driven by many-body correlations. 
The Hamiltonian in Eq.~\eqref{eq:ham_massb} is integrable and a complete solution could be obtained using the Bethe ansatz~\cite{Pehlivan2011}. This was used in Ref.~\cite{Cervia2019} to study a time-dependent generalization of the model for small systems.

At this point it is convenient to introduce the total spin operator $\vec{J}=\sum_i\vec{\sigma}_i/2$ and correspondingly $\vec{J}^A$ and $\vec{J}^B$ for the neutrinos in the two beams. The full Hamiltonian in Eq.~\eqref{eq:ham_massb} commutes with both the $z$ component of the total spin, $J_z$, and its magnitude, $J_z^2$. As the initial state $\ket{\Psi_0}$ is an eigenstate of $J_z$, we can rewrite the relevant Hamiltonian for our setup as follows
\begin{equation}
\label{eq:spinham}
H = -\delta_\omega \left(J_z^A-J_z^B\right) + \frac{\mu}{N} J^2\;,
\end{equation}
where we dropped irrelevant constant factors and introduced a single parameter $\delta_\omega=(\omega_A-\omega_B)/2$ for the energy difference between the two beams. This model is a variant of the Lipkin-Meshov-Glick (LMG) model~\cite{Lipkin1965} explored extensively in the past~\cite{Vidal2004,Vidal2004b,Vidal2004c,Latorre2005,Ribeiro2008}. The model with no vacuum term merits special attention as it is diagonal in the angular momentum basis and exact results for the flavor dynamics can be obtained as shown in Refs.~\cite{Friedland2003,Friedland2006}. In this model, collective flavor oscillation are shown to occur at time scales much longer than $\tau_F$ with the scaling $\tau_L=\mu^{-1}\sqrt{N}$ expected from incoherent scattering from the background neutrinos~\cite{Friedland2003,Friedland2003b}. Due to this strong scaling with system size, when $N\gg1$ the flavor dynamics is essentially frozen and the mean field prediction of no evolution eventually becomes exact. This has been taken as an indication that the mean field description is appropriate for large systems and that entanglement does not play any major role~\cite{Friedland2003b,Friedland2003}.

As a way to track flavor oscillations, we compute the survival probability $p(t)=(1-\langle\sigma^z_1(t)\rangle)/2$ for a neutrino (we take the first) to be found in it initial flavor state (in this case $\ket{\downarrow}$). By using the exchange symmetry for neutrinos within each beam of both the Hamiltonian and the initial state $\ket{\Psi_0}$, together with energy conservation, we can write the expectation value of the total spin as
\begin{equation}
\label{eq:jsquared}
\langle J^2(t)\rangle = \frac{N}{2}\left(1+2\frac{\delta_\omega}{\mu} N \left(1-p(t)\right)\right)\;.
\end{equation}
When $\delta_\omega/\mu<0$ the total spin, starting already at a relatively small value, can only decrease further during time evolution. Since $J^2$ is positive semi-definite, this introduces a lower bound $p(t)\geq1-\mu/(2N\delta_\omega)$ and the neutrinos remain frozen in the initial state in the thermodynamic limit.
In the opposite limit, for $\delta_\omega/\mu \geq 1/4$ flavor oscillations become parametrically small as $p(t)\geq1-\mu/(4\delta_\omega)$ but remain finite in the $N\gg1$ limit.
In the intermediate regime $0<\delta_\omega/\mu\lesssim 1/4$ we instead expect to find large amplitude flavor oscillations. 

We solve for the real-time evolution of the system using a Matrix Product State(MPS) ansatz for the many-body wave function~\cite{Vidal2003_mps,Vidal_2004_mps,Paeckel2019}. The approach has a controllable error and it's computationally efficient for systems with low levels of bipartite entanglement~\cite{Vidal2003_mps,Schuch2008} and is described in detail in the next Section.

\section{Matrix Product State Simulation}
\label{sec:MPS}
In this section we review the Matrix Product State (MPS) technique used to generate the results in this work. Additional information can be found in the original paper~\cite{Vidal2003_mps} and in more recent reviews~\cite{Schollwock2011,Paeckel2019}. In order to implement long-range interactions efficiently with a MPS, we use the same strategy introduced recently for circuit based simulations in~\cite{Hall2021}, a brief explanation of the construction is presented in the last subsection. In the last section we also compare directly the mean-field approximation to an MPS-based simulation with unit bond dimension.

\subsection{MPS Representation}
Consider a general quantum state $\ket{\Psi}$ of a system with $N$ spin-$1/2$ particle, it's wavefunction can be expressed as a $N$-component tensor as
\begin{equation}
\ket{\Psi} = \sum_{\sigma_1,\dots,\sigma_N} \Psi^{\sigma_1\cdots\sigma_N}\ket{\sigma_1\cdots\sigma_N}\;,
\end{equation}
with $\ket{\sigma_1\cdots\sigma_N}$ one of the $2^N$ basis states spanning the total Hilbert space of the system. A Matrix Product State is an exact representation of the $N$-component tensor $\Psi^{\sigma_1\cdots\sigma_N}$ as a product of $N$, smaller, $3$-component tensors $A^{\sigma_k}_{\alpha,\beta}$. A constructive procedure to find this representation, starting from the original tensor $\Psi^{\sigma_1\cdots\sigma_N}$, can be obtained by applying $N$ times a Singular Value Decomposition (SVD) of appropriately reshaped tensors. This decomposition allows to represent an arbitrary rectangular matrix $M$ of dimension $D_A\times D_B$ as the following product of three matrices
\begin{equation}
\label{eq:app_svd}
M_{ij} = \sum_k U_{ik} \Sigma_{kk} V^{\dagger}_{kj}\;,
\end{equation}
where $U$ is a $D_A\times\min(D_A,D_B)$ matrix with orthonormal columns, $\Sigma$ is a diagonal matrix of dimension $\min(D_A,D_B)\times\min(D_A,D_B)$ with either zero or positive entries and $V^\dagger$ is a $\min(D_A,D_B)\times D_B$ matrix with orthonormal rows. An important property of the SVD, which we will use repeatedly in the derivation below, is that the approximation of the original matrix $M$, with rank $r\leq\min(D_A,D_B)$, using a second matrix $M'$ with rank $r'<r$ can be done optimally, in the Frobenius norm, by truncating the sum in Eq.~\eqref{eq:app_svd} so that we keep only the $r'$ largest singular values $\Sigma_{kk}$. We will assume that the SVD is performed with the convention that singular values in $\Sigma$ are ordered as $\Sigma_{11}\geq\cdots\geq\Sigma_{NN}$. With this convention, the optimal approximation corresponds to truncating the sum as
\begin{equation}
M'_{ij} = \sum_k^{r'} U_{ik} \Sigma_{kk} V^{\dagger}_{kj}\;.
\end{equation}

Starting from the first spin on the left, we first reshape $\Psi^{\sigma_1\cdots\sigma_N}$ into the rectangular matrix $\Psi_{\sigma_1(\sigma_2\cdots\sigma_N)}$ of dimension $2\times2^{N-1}$. An explicit SVD of this matrix using Eq.~\eqref{eq:app_svd} will then produce the following decomposition
\begin{equation}
\begin{split}
\Psi_{\sigma_1(\sigma_2\cdots\sigma_N)} &= \sum_{\alpha_1}^{r_1} U_{\sigma_1\alpha_1}\Sigma_{\alpha_1\alpha_1} V^\dagger_{\alpha_1(\sigma_2\cdots\sigma_N)}\\
&=\sum_{\alpha_1}^{r_1} U_{\sigma_1\alpha_1} W_{\alpha_1\sigma_2\cdots\sigma_N}\\
&=\sum_{\alpha_1}^{r_1} A^{[1]\sigma_1}_{\alpha_1} \Psi^{[1]}_{(\alpha_1\sigma_2)(\sigma_3\cdots\sigma_N)}
\;,
\end{split}
\end{equation}
with $r_1\leq2$ the rank of the original rectangular matrix and in the second line we have reshaped the product of $\Sigma$ and $V^\dagger$ to a vector $\vec{W}$. In the last line, we have further reshaped the matrix $U$ into two row vectors, with components $A^{[1]\sigma_1}_{\alpha_1}=U_{\sigma_1\alpha_1}$, and the vector $\vec{W}$ to a $2r_1\times2^{N-2}$ matrix $\Psi^{[1]}_{(\alpha_1\sigma_2)(\sigma_3\cdots\sigma_N)}$. At this point, we can repeat the procedure by applying a SVD to this last matrix and proceed from left to right until we have reached the last spin of the chain. The resulting decomposition becomes
\begin{equation}
\begin{split}
\label{eq:mps_left}
\Psi^{\sigma_1\cdots\sigma_N} &= \sum_{\alpha_1,\dots,\alpha_{N-1}} A^{[1]\sigma_1}_{\alpha_1}A^{[2]\sigma_2}_{\alpha_1\alpha_2}\cdots A^{[N-1]\sigma_{N-1}}_{\alpha_{N-2}\alpha_{N-1}}A^{[N]\sigma_N}_{\alpha_{N-1}}\\
&= \mathbf{A}^{[1]\sigma_1}\mathbf{A}^{[2]\sigma_2}\cdots \mathbf{A}^{[N-1]\sigma_{N-1}}\mathbf{A}^{[N]\sigma_N}\;,
\end{split}
\end{equation}
where in the last line we suppressed the lower indices and replaced them by matrix multiplications. Using this construction, a general state $\ket{\Psi}$ can be written as a Matrix Product State. Note that the rank of the SVD in the center of the chain could be as large as $2^{N/2-1}$ (assuming $N$ even) and storage requirements for representing a general state will then scale exponentially with the system size. 

As we will see soon, the maximum rank $r_{max}=\max_k r_k$ needed to describe accurately a given state is related to the maximum bipartite entanglement entropy in the system. In order to see this, it is first useful to introduce a different construction for the MPS. Consider a similar iterative decomposition procedure for the tensor $\Psi^{\sigma_1\cdots\sigma_N}$ as the one described above, but now starting from the right-most site. The first step leads to
\begin{equation}
\begin{split}
\Psi_{(\sigma_1\cdots\sigma_{N-1})\sigma_N} &= \sum_{\beta_N}^{r_N} U_{(\sigma_1\cdots\sigma_{N-1})\beta_N}\Sigma_{\beta_N\beta_N} V^\dagger_{\beta_N\sigma_N}\\
&=\sum_{\beta_N}^{r_N} W'_{\sigma_1\cdots\sigma_{N-1}\beta_N}V^\dagger_{\beta_N\sigma_N}\\
&=\sum_{\beta_N}^{r_N}\Psi^{[N]}_{(\sigma_1\cdots\sigma_{N-2})(\sigma_{N-1}\beta_N)} B^{[N]\sigma_N}_{\beta_N}\;,
\end{split}
\end{equation}
where we have introduced the two column vectors $B^{[N]\sigma_N}_{\beta_N}=V^\dagger_{\beta_N\sigma_N}$ and the residual matrix has maximum dimensions given by $2^{N-2}\times2r_N$. Proceeding with the same procedure until the left edge, we find another matrix product state with amplitudes
\begin{equation}
\begin{split}
\label{eq:mps_right}
\Psi^{\sigma_1\cdots\sigma_N} &= \mathbf{B}^{[1]\sigma_1}\mathbf{B}^{[2]\sigma_2}\cdots \mathbf{B}^{[N-1]\sigma_{N-1}}\mathbf{B}^{[N]\sigma_N}\;.
\end{split}
\end{equation}
The two representations are related by a gauge transformation~\cite{Schollwock2011,Paeckel2019}, and they are characterized by different normalization properties. In particular, we have
\begin{equation}
\label{eq:norm_tens}
\sum_{\sigma_k} \mathbf{A}^{[k]\sigma_k\dagger}\mathbf{A}^{[k]\sigma_k}=\mathbb{1}\quad\sum_{\sigma_k} \mathbf{B}^{[k]\sigma_k}\mathbf{B}^{[k]\sigma_k\dagger}=\mathbb{1}\;.
\end{equation}
For this reason, the state resulting from the decomposition in Eq.~\eqref{eq:mps_left} in terms of $\mathbf{A}$ tensors is called a {\it left-canonical} MPS, while the construction in terms of $\mathbf{B}$ tensors in Eq.~\eqref{eq:mps_right} is called {\it right-canonical}~\cite{Schollwock2011,Paeckel2019}.

In order to see more clearly the connection between entanglement and the maximum rank in the decomposition, also called {\it bond dimension}, it is useful to introduce yet another canonical form.  This is obtained by starting with a decomposition from the left up to a chosen site $k$, at this point we start with a decomposition from the right and we stop when we reach site $k+1$. We can write the result of this decomposition as 
\begin{equation}
\begin{split}
\label{eq:mps_sitec}
\Psi^{\sigma_1\cdots\sigma_N} &= \mathbf{A}^{[1]\sigma_1}\cdots\mathbf{A}^{[k]\sigma_k}\mathbf{\Sigma}^{[k]} \mathbf{B}^{[k+1]\sigma_{k+1}}\cdots\mathbf{B}^{[N]\sigma_N}\;,
\end{split}
\end{equation}
with the diagonal matrix $\mathbf{\Sigma}^{[k]}$ containing the singular values at the bond between the $k$ and the $k+1$ sites. The site $k$ is called the {\it orthogonality center}, since tensors to the left are left-normalized while tensors on the right are right-normalized. This property allows us to write the state in this representation as
\begin{equation}
\label{eq:bipart_state}
\ket{\Psi} = \sum_{\alpha_k}^{r_k} \Sigma^{[k]}_{{\alpha_k}{\alpha_k}} \ket{\Phi^L_{\alpha_k}}\otimes\ket{\Phi^R_{\alpha_k}}\;,
\end{equation}
where we have defined two sets of states
\begin{equation}
\ket{\Phi^L_{\alpha_k}} = \sum_{\sigma_1,\dots,\sigma_k} \left[\mathbf{A}^{[1]\sigma_1}\cdots\mathbf{A}^{[k]\sigma_k}\right]_{\alpha_k} \ket{\sigma_1\cdots\sigma_k}\;,
\end{equation}
and 
\begin{equation}
\ket{\Phi^R_{\alpha_k}} = \sum_{\sigma_{k+1},\dots,\sigma_N} \left[\mathbf{B}^{[k+1]\sigma_{k+1}}\cdots\mathbf{B}^{[N]\sigma_N}\right]_{\alpha_k} \ket{\sigma_{k+1}\cdots\sigma_N}\;.
\end{equation}
Thanks to the normalization properties of the $\mathbf{A}$ and $\mathbf{B}$ tensors from Eq.~\eqref{eq:norm_tens}, we see that these states form orthonormal sets of states for the left and right Hilbert spaces respectively. We can then interpret Eq.~\eqref{eq:bipart_state} as the Schmidt decomposition of the state $\ket{\Psi}$ on the left-right bipartition. The number of states in the sum can then be identified with the {\it Schmidt rank}, which is a natural measure of entanglement between states on the left partition and states on the right one. 

In particular, the Von-Neumann entanglement entropy for the left-right bipartition at site $k$ of the state $\ket{\Psi}$ can be expressed as a function of the singular values as
\begin{equation}
S_k = -\sum_{\alpha_k=1}^{r_k} \Sigma^{[k]2}_{{\alpha_k}{\alpha_k}}\log_2\left(\Sigma^{[k]2}_{{\alpha_k}{\alpha_k}}\right)\;.
\end{equation}
The {\it Schmidt rank} $r_k$ gives the number of non-zero singular values so that the entanglement is bounded by
\begin{equation}
S_k\leq \log_2(r_k)\;.
\end{equation}
We can now see that MPS with small bond dimensions correspond to quantum states with little entanglement~\cite{Vidal2003_mps}. Before moving on to the time-dependent simulation, it's important to mention that the {\it mixed-canonical} form in Eq.~\eqref{eq:mps_sitec} is also particularly convenient to find more efficient approximations to an initial MPS. Given an MPS $\ket{\Psi}$ with bond dimension $r_k$ at the bond between site $k$ and $k+1$, we can find another MPS $\ket{\Psi'}$ with lower bond dimension $r'_k<r_k$ so that the error in $2$-norm is bounded from above by~\cite{Verstraete2006}
\begin{equation}
\left\|\ket{\Psi}-\ket{\Psi'}\right\|^2_2 \leq 2 \sum_{\alpha_k=r_k'+1}^{r_k} \Sigma^{[k]2}_{{\alpha_k}{\alpha_k}}\;.
\end{equation}
Thanks to the bi-orthogonality of the Schmidt decomposition Eq.~\eqref{eq:bipart_state}, we can obtain $\ket{\Psi'}$ from $\ket{\Psi}$ by keeping only the $r_k'$ largest singular values in the sum. If the singular values decrease sufficiently fast, a good approximation to the original state could be obtained with an MPS with a much smaller bond dimension~\cite{Vidal2003_mps}.

\subsection{Time evolution of a MPS}
\label{sec:mps_time_evol}
We now turn to explain how one can simulate the time-evolution for the general neutrino Hamiltonian in Eq.(1) of the main text. The basic operation we will use repeatedly in the following is the application of an $SU(4)$ operation on a pair of nearest neighbor indices $(k,k+1)$. A generic $4\times 4$ unitary $U$ of this type can be represented as a $4$-component tensor $G^{\sigma_1\sigma_2}_{\sigma_1'\sigma_2'}$ with each index having dimension equal to $2$. When we contract this tensor with the MPS, only the tensors around the bond $(k,k+1)$ get transformed. More specifically, we have that the matrices
\begin{equation}
\mathbf{C}^{\sigma_k\sigma_{k+1}} = \mathbf{A}^{[k]\sigma_k}\mathbf{\Sigma}^{[k]} \mathbf{B}^{[k+1]\sigma_{k+1}}\;,
\end{equation}
get updated into
\begin{equation}
\label{eq:updated_t}
\begin{split}
\widetilde{\mathbf{C}}^{\sigma_k\sigma_{k+1}} &= \sum_{\sigma_k'\sigma_{k+1}'}G^{\sigma_k\sigma_{k+1}}_{\sigma_k'\sigma_{k+1}'}\mathbf{A}^{[k]\sigma_k'}\mathbf{\Sigma}^{[k]} \mathbf{B}^{[k+1]\sigma_{k+1}'}\;.
\end{split}
\end{equation}
We can now restore the MPS form by performing one final SVD of $\widetilde{\mathbf{C}}^{\sigma_k\sigma_{k+1}}$ properly reshaped as a matrix
\begin{equation}
\widetilde{C}_{(\alpha_{k-1}\sigma_k)(\sigma_{k+1}\beta_{k+2})}= \sum_{\alpha_k}^{\widetilde{r}_k} \widetilde{U}_{(\alpha_{k-1}\sigma_k)\alpha_k}\widetilde{\Sigma}_{\alpha_k\alpha_k} \widetilde{V}^\dagger_{\alpha_k(\sigma_{k+1}\beta_{k+2})}\;,
\end{equation}
with a possibly larger bond dimension $\widetilde{r}_k$. Now, by identifying $\widetilde{\mathbf{A}}^{[k]\sigma_k}$ with $\widetilde{U}$, $\widetilde{\mathbf{B}}^{[k+1]\sigma_{k+1}}$ with $\widetilde{V}^\dagger$ and the singular vector matrices $\widetilde{\mathbf{\Sigma}}^{[k]}$ with $\widetilde{\Sigma}$, we have a new MPS with decomposition
\begin{equation}
\begin{split}
\label{eq:mps_sitec_updt}
\widetilde{\Psi}^{\sigma_1\cdots\sigma_N} &= \mathbf{A}^{[1]\sigma_1}\cdots\widetilde{\mathbf{A}}^{[k]\sigma_k}\widetilde{\mathbf{\Sigma}}^{[k]} \widetilde{\mathbf{B}}^{[k+1]\sigma_{k+1}}\cdots\mathbf{B}^{[N]\sigma_N}\;.
\end{split}
\end{equation}
The SVD requires $\mathcal{O}((r_{k-1}+r_{k+1})r^2_{k+1})$ operations and is efficient only if we are able to keep a small bond dimension by truncating the MPS as shown above.

\begin{figure*}[bt]
 \centering
 \includegraphics[width=0.98\textwidth]{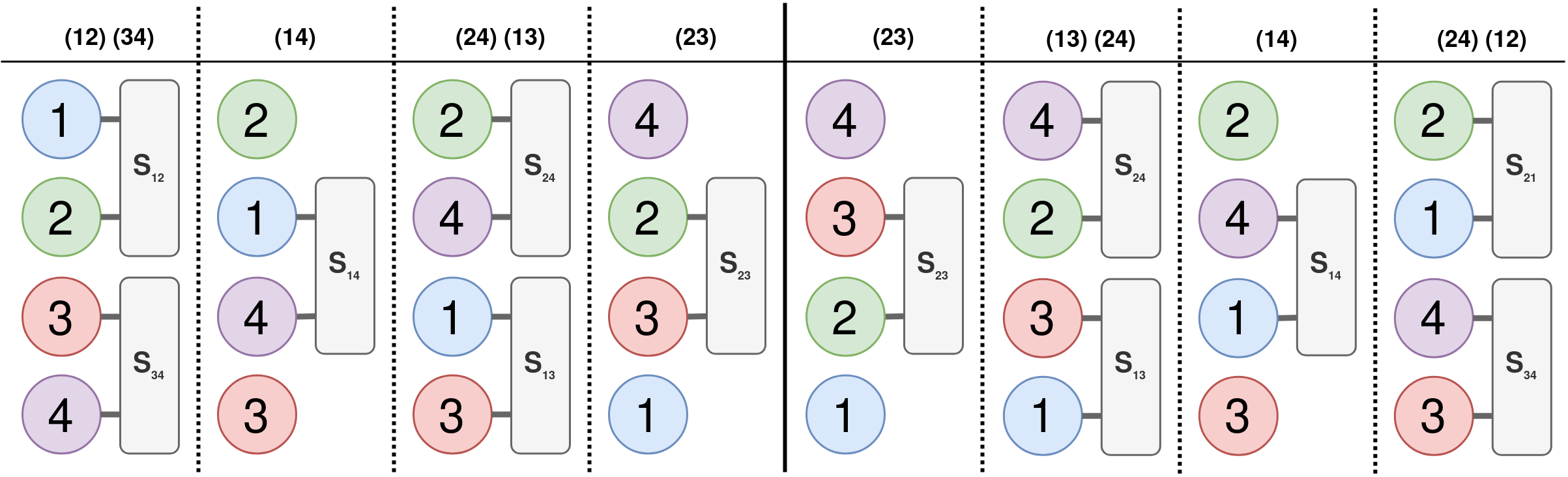}
 \caption{(Color online) Schematic representation of one step of time evolution for a system with $N=4$ neutrinos. The label on the top indicate the Hamiltonian terms $h_{ij}$ considered at each one of the $2N=8$ sub-steps.}
\label{fig:swap_network}
\end{figure*} 

With nearest-neighbor two-site operators at our disposal, we can now present the scheme used to simulate time evolution under Hamiltonians of the form
\begin{equation}
\label{eq:sm_ham}
H = \sum_{k=1}^N \vec{B}_{k}\cdot\vec{\sigma}_k + \sum_{i<j}^N \mathcal{V}_{ij}\vec{\sigma}_i\cdot\vec{\sigma}_j\;,
\end{equation}
with arbitrary long range interactions. For convenience we first rewrite the Hamiltonian as a sum of, not necessarily nearest-neighbor, pair Hamiltonians $h_{ij}$ as
\begin{equation}
\label{eq:app_ham}
H = \sum_{i<j}^N\left[\frac{\vec{B}_{i}\cdot\vec{\sigma}_i+\vec{B}_{j}\cdot\vec{\sigma}_j}{N-1}+\mathcal{V}_{ij}\vec{\sigma}_i\cdot\vec{\sigma}_j\right] = \sum_{i<j}^Nh_{ij}.
\end{equation}
The time evolution operator under this Hamiltonian can be approximated for short times using
\begin{equation}
\label{eq:time_approx}
e^{-i\delta H} = \prod_{i<j=1}^Ne^{-i\delta h_{ij}} \prod^{1}_{i<j=N} e^{-i\delta h_{ij}} + \mathcal{O}\left(\delta^3\|H\|^3\right)\;,
\end{equation}
where the second sequence of products are performed in reverse order. This is the standard second-order Trotter-Suzuki approximation for the time evolution operator~\cite{Suzuki91}. Each one of the operators appearing in Eq.~\eqref{eq:time_approx} is a $SU(4)$ unitary acting on all pairs of sites for a total of $N(N-1)$ gates. In order to replace long range gates into nearest-neighbor ones, it is convenient to introduce the SWAP operation. This is defined as
\begin{equation}
\text{SWAP} = \begin{pmatrix}
1&0&0&0\\
0&0&1&0\\
0&1&0&0\\
0&0&0&1\\
\end{pmatrix}\;.
\end{equation}
When applied to a pair of sites, it exchanges the state as
\begin{equation}
\text{SWAP}\ket{\phi}\otimes\ket{\psi} = \ket{\psi}\otimes\ket{\phi}\;.
\end{equation}
Using this operation we can transport two sites next to each other, apply the unitary $T_{ij}=\exp(-i\delta h_{ij})$ as a nearest-neighbor operation and finally swap the two sites back to their original position~\cite{Stoudenmire2010}. Given the all-to-all nature of the pair interaction in Eq.~\eqref{eq:app_ham}, a naive application of this strategy will result in $\mathcal{O}(N^3)$ nearest-neighbor operations to implement one step of time evolution using the approximation in Eq.~\eqref{eq:time_approx} for the propagator.

In this work we employ instead the swap-network scheme introduced in Ref.~\cite{Hall2021} for studying neutrino dynamics on a digital quantum computer with linear connectivity. The construction is inspired by earlier work on fermionic swap-networks~\cite{Kivlichan2018} and relies on the observation that, if we perform $N$ layers of nearest-neighbor swap operations by alternating the even bonds with the odd bonds, we can bring every site next to every other site at least once using a total of $N(N-1)/2$ operations. At the end of these $N$ steps, the sites will be left in reverse order. It is now clear that, if we define a modified propagator $S_{ij}$ as follows (with $h_{ij}$ from Eq.~\eqref{eq:app_ham})
\begin{equation}
S_{ij} = \text{SWAP}\times\exp(-i\delta h_{ij})\;,
\end{equation}
and apply nearest-neighbor $S_{ij}$ operations instead of swaps as before, at the end of the sequence we will have implemented the first product of pair operators in the approximation Eq.~\eqref{eq:time_approx} for the propagator. If we now apply the sequence in reverse order we can complete a full time step and obtain a state where sites have the original order.  One can express this more directly by explicitly  denoting the sites where the unitary operation is applied as superscript indices $S^{a,b}_{i,j}$. With this notation, and assuming $N$ even for simplicity, we can express the total unitary $Q^e_k$ on the $k$-th even layer and $Q^o_k$ on the $k$-th odd layer as follows
\begin{equation}
\begin{split}
Q^e_k &= \prod_{a=1}^{N/2} S^{2a-1,2a}_{i(2a-1,k,e),j(2a,k,e)} \\
Q^o_k &= \prod_{a=1}^{N/2-1} S^{2a,2a+1}_{i(2a,k,o),j(2a+1,k,o)}\;,
\end{split}
\end{equation}
where $i(a,k,e/o)$ and $j(a,k,e/o)$ denote the amplitude residing on site $a$ at the $k$-th even or odd step. The mapping between sites and amplitudes evolves as the algorithm proceeds but it is straightforward to keep track of this and assign the correct amplitude index at every point in the step. The full second-order step from Eq.~\eqref{eq:time_approx} can then be expressed as
\begin{equation}
\prod_{i<j=1}^Ne^{-i\delta h_{ij}} \prod^{1}_{i<j=N} e^{-i\delta h_{ij}} = \prod_{k=1}^{N/2} \left[Q^e_k Q^o_k\right] \prod_{k=N/2}^{1} \left[Q^o_k Q^e_k\right]\;.
\end{equation}

A schematic depiction of a complete step for a system with $N=4$ neutrinos is displayed in Fig.~\ref{fig:swap_network}. In order to avoid cumbersome notation, in the figure only the lower indices $(ij)$ in the propagators $S_{ij}$ are used and these identify the neutrino amplitude associated with the specific sites the propagator is acting upon at any given step.

In this work we perform a SVD truncation to MPS form after each application of the nearest neighbor operation $S_{ij}$ using a threshold of $10^{-10}$ for the discarded eigenvalues. Better truncation schemes are possible (see eg.~\cite{Paeckel2019}) and their potential advantages will be explored in future work.

Finally, we decompose the full time evolution for a total time $t$ into steps of size $\delta\ll t$ as
\begin{equation}
\label{eq:full_prop}
\exp\left(-it H\right) = \left[\exp\left(-i\delta H\right)\right]^{t/\delta}\;.
\end{equation}
We then approximate each short-time propagator for time $\delta$ using the swap-network scheme described above.

For all the results presented in the main text, we used a fixed time-step $\delta=0.05\mu^{-1}$ which we found provided a good compromise between precision and efficiency for the systems studied here (see also Appendix~\ref{app:tstep} for a more complete discussion). As already mentioned above, the overall scheme remains efficient provided the singular values on each bond decay sufficiently fast to zero that a good approximation to the MPS can be obtained using a low bond dimension. Thanks to the (at most) logarithmic scaling of the entanglement entropy observed in the systems studied in main text, the maximum bond dimension required to keep the truncation error below $10^{-10}$ was $\approx N/2$ allowing us to easily explore systems with $N\gtrapprox100$. The software developed to carry out the simulations reported in this work used the C++ version of the ITensor library which provides efficient representations for MPS and more general tensors~\cite{itensor}.

\section{Results}
\label{sec:res}

\begin{figure}[b]
 \centering
 \includegraphics[width=0.45\textwidth]{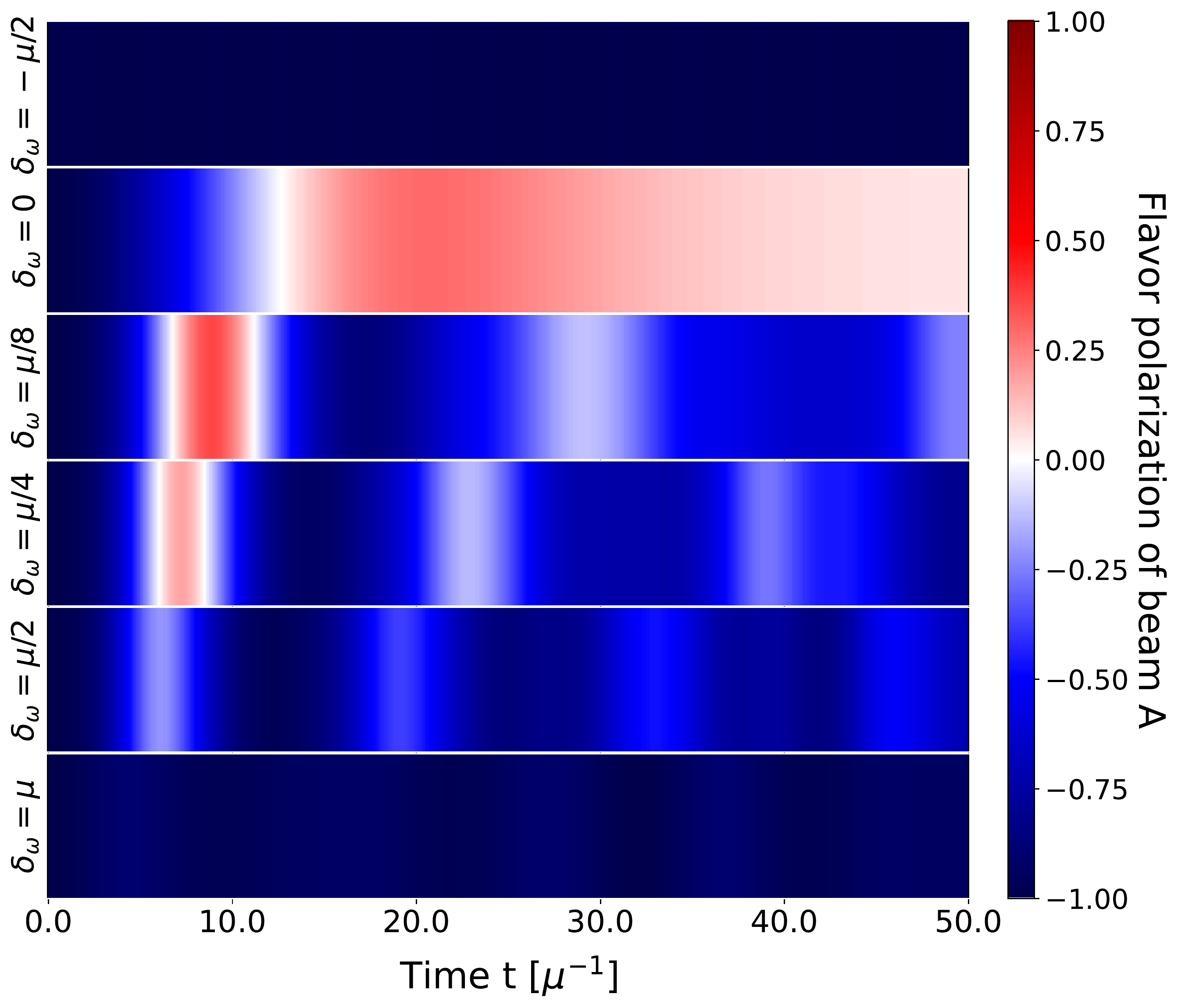}
 \caption{(Color online) Flavor polarization per particle $\langle J^A_z(t)\rangle/(N/4)$ of neutrinos in the A beam as a function of time for six values of the energy asymmetry parameter $\delta_\omega/\mu$ (from top to bottom): $-0.5,0.0,0.125,0.25,0.5,1.0$. }
\label{fig:tev}
\end{figure}

\begin{figure}[t]
 \centering
 \includegraphics[width=0.49\textwidth]{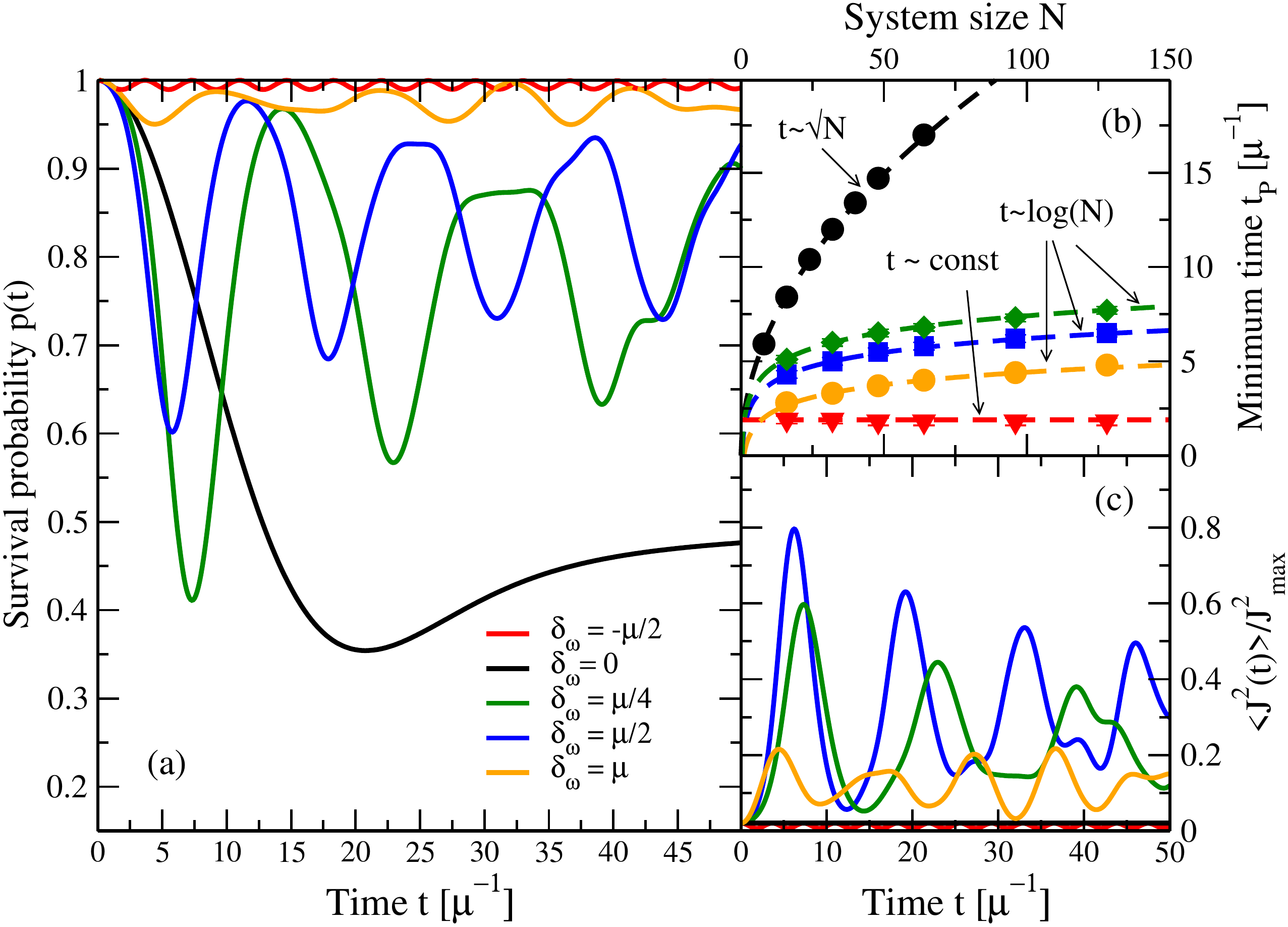}
 \caption{(Color online) Panel (a) shows the survival probability $p(t)$ in a system of $N=96$ neutrinos  initialized in $\ket{\Psi_0}$ and evolved with different values of the one body asymmetry parameter. Panel (b) shows the evolution of the minimum time $t_P$ as a function of system size and panel (c) shows the (normalized) expectation value of the total angular momentum for different values of $\delta_\omega$ in a system of $N=96$ neutrinos.}
\label{fig:1b_pers}
\end{figure}

The qualitative difference on the collective flavor oscillations in the different dynamical phases can be observed directly from the time evolution of the flavor polarization. Fig.~\ref{fig:tev} shows the average flavor polarization of the neutrinos in the $A$ beam $\langle J^A_z(t)\rangle/(N/4)=1-2p(t)$, for a system of $N=96$ neutrinos and different values for the asymmetry energy $\delta_\omega$. Due to the conservation of $J_z$, the polarization of beam B is simply the inverse. We can see strong flavor oscillations for $0<\delta_\omega\lesssim\mu$ which for $\delta_\omega<\mu/2$ can even lead to a net flavor inversion in the beam (red areas in Fig.~\ref{fig:tev}). In the stable phases for large values of $\delta_\omega/\mu$ (bottom panel) or negative asymmetries (top panel), the amplitude of these oscillations is greatly reduced. More detail on these flavor oscillations is given in panel (a) of Fig.~\ref{fig:1b_pers} where we show the results for the survival probability $p(t)$ and the same system of $N=96$ neutrinos. For negative values of $\delta_\omega$ the polarization experiences small oscillations around the initial state, as expected from our the discussion on the evolution of the total angular momentum based on Eq.~\eqref{eq:jsquared}.

In the phase with $0<\delta_\omega\lesssim\mu$ the survival probability experiences instead fast oscillations on time scales much shorter than in the high density limit $\delta_\omega =0$. If, in the latter case, the first minimum of the survival probability is reached for a time $t_P$ that scales with $\tau_L\propto\sqrt{N}$, in the former case we have instead evolution at the fast scale $\tau_F\propto\log(N)$ found in Refs.~\cite{Bell2003,sawyer2004classical}, but now for a physical model with $SU(2)$ invariance (see also~\cite{Roggero2021b}). The difference in scaling of $t_P$ with system size is evident from the results presented in panel (b) of Fig.~\ref{fig:1b_pers}. For all values of $\delta_\omega\neq0$ a good fit to data is obtained with the ansatz $\mu t_P(N) = a_t \log(N) + c_t$. The best fit values for the parameters extracted from our simulations are reported in Tab.~\ref{tab:oneb_times}, together with the extrapolated value of the minimum of the survival probability $p_{min}$ in the $N\gg1$ limit. Even though it doesn't directly quantify the final flavor content in the long time limit, this quantity is still remarkably useful as it provides an upperbound on the maximum amplitude of flavor oscillations. In particular we see that, for $\delta_\omega$ outside the critical region $(0.0,1.0)$, the latter converges to $p_{min}\approx1$ in the thermodynamic limit and no flavor evolution is present. Interestingly, the fluctuation timescales in the unstable region seem to scale approximately as $a_t\propto \sqrt{\mu/\delta_\omega}$ similarly to the more familiar bipolar oscillations~\cite{Kostelecky1995,Hannestad2006,Duan2007b}. The instability observed in our simulations is also similar to bipolar oscillations in that, if we take the beam $B$ to be composed of anti-neutrinos (with negative $\omega_B$~\cite{Duan2010} in normal hierarchy) and beam $A$ of neutrinos, the appearance of the instability leading to oscillations in our setting depends on the neutrino hierarchy: for inverted hierarchy, the single particle energies change sign and the instability cannot occur because $\delta_\omega<0$ irrespective of the energies, for normal hierarchy we have always positive energy differences and oscillations can occur when $\omega_A+|\omega_B|\lesssim\mu$. The situation is reversed exchanging neutrinos with anti-neutrinos. Finally, fast oscillations can also occur if both beams are composed by neutrinos (or anti-neutrinos) provided that
\begin{equation}
\label{eq:inst_cond}
0\leq\pm\frac{E_B-E_A}{E_AE_B}\lesssim \frac{4\sqrt{2}G_F\rho_\nu}{\Delta m^2}\;,
\end{equation}
with $+$ for $\nu$ and $-$ for $\bar{\nu}$ in the normal hierarchy~\footnote{Note that this condition can be modified by the presence of an MSW-like matter term which couple differently to the two flavors.}.

\begin{table}[t]
\begin{tabular}{c|c|c|c|c}
$\delta_\omega/\mu$ & $a_t$ & $b_t$& $c_t$ &$p_{min}$\\ \hline
-0.5 & 0 & 0 & 1.8(2)& 1.00(1)\\
0.0 & 0 & 2.10(5) & 0&0.36(1)\\
0.01 & 5.4(2) & 0 & -8(1)& 0.35(2) \\
0.05 & 2.58(6) & 0 & 0 & 0.31(4)\\
0.125 & 1.65(6)& 0& 1.4(2)& 0.32(3) \\
0.25 & 1.22(4) & 0 & 1.6(2) & 0.41(3)\\
0.5 & 1.06(4) & 0 & 1.4(2) & 0.61(4)\\
1.0 & 0.96(5) & 0 &0 & 0.99(2) 
\end{tabular}
\caption{Optimal parameters of the model discussed in the main text $\mu t_P(N) = a_t \log(N) + b_t\sqrt{N} +c_t$ for time to reach the minimum of the survival probability $t_P$. The minimum of the survival probability in the thermodynamic limit is extracted using the ansatz $p_{min}(N) = p_{min} + p_c/N^\gamma$ with exponents $\gamma=1$ for $\delta_\omega=0$ and $\gamma=0.5$ otherwise.}
\label{tab:oneb_times}
\end{table}

In panel (c) of Fig.~\ref{fig:1b_pers} we show the, normalized, expectation value of the total angular momentum measured using the relation in Eq.~\eqref{eq:jsquared} derived above. As we can see from the figure, in the unstable region the system can acquire a considerable fraction of the maximum possible total angular momentum, especially for $\delta_\omega=\mu/2$ (blue line) where it reaches a maximum of $\approx80\%$.

As an alternative way to quantify quantum correlations in the evolved state, we compute the half-chain entanglement entropy (see eg.~\cite{Eisert2010}) defined explicitly as 
\begin{equation}
\label{eq:entropy}
S_{N/2}(t) = -Tr\left[\rho_B(t)\log_2\left(\rho_B(t)\right)\right]\;.
\end{equation}
Here the density matrix $\rho_B=Tr_A\left[\rho(t)\right]$ is obtained by tracing the full density matrix $\rho(t)=\rvert\Psi(t)\rangle\langle\Psi(t)\lvert$ of the system at time $t$ over the first $N/2$ neutrinos.

\begin{figure}[t]
 \centering
 \includegraphics[width=0.49\textwidth]{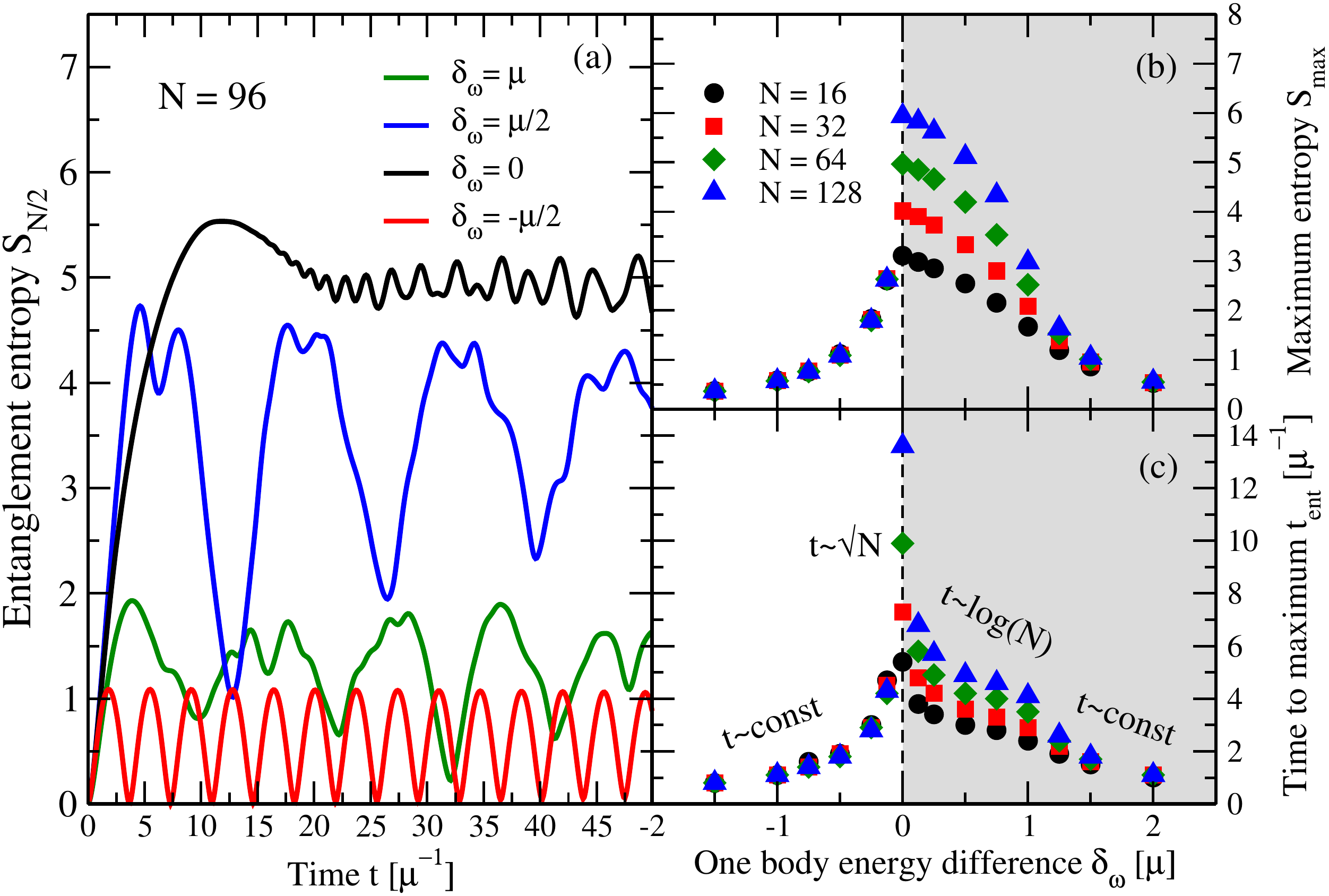}
 \caption{(Color online) Half chain entanglement entropy $S_{N/2}$ for $N=96$ neutrinos and different values of the one-body energy difference: the orange line is $\delta_\omega= \mu$, the blue line is for $\delta_\omega=\mu/2$, the red line for $\delta_\omega=-\mu/2$ and the black line has $\delta_\omega=0$. Panel (b) shows the maximum entropy $S_{max}$ for three system sizes $N=16,32,64,128$ and different values of $\delta_\omega$. Panel (c) shows the time $t_{ent}$ to reach the maximum. }
\label{fig:1b_entropy_wt}
\end{figure}

The presence of different dynamical phases can be seen directly from the results on the half-chain entanglement entropy $S_{N/2}(t)$ presented in Fig.~\ref{fig:1b_entropy_wt}. Panel (b) shows the maximum value $S_{max}$ that is reached by the entanglement entropy for different values of $\delta_\omega$ and system sizes while in panel (a) we show examples of the time evolution of $S_{N/2}(t)$ for a system of $N=96$ neutrinos.
In the phase for $\delta_\omega<0$, the behavior is similar to a gapped system with $S_{max}$ independent of system size while we find $S_{N/2}(t)$ to oscillate in time (red curve) with an approximately constant frequency.
We recover a similar result for $\delta_\omega\gtrsim\mu$ when, as expected from the discussion above, the system becomes again approximately frozen in the initial configuration. As shown in panel (a) for the case $\delta_\omega=\mu$ (orange line), the oscillations in the entanglement entropy are much less regular than in the $\delta_\omega<0$ phase and do not return close to zero.
At the transition point $\delta_\omega=0$ we find that, after an initial rapid growth phase, the entropy $S_{N/2}(t)$ reaches a peak at $S_{max}\approx\log_2(N/2)$ and then plateaus at a slightly smaller value with oscillations around the average. This maximum value for the entanglement entropy is reminiscent to the one in ground states of spin systems at a quantum critical point~\cite{Vidal2003,Refael2004} and reflects the absence of a gap in the Hamiltonian. This connection allows to understand why it is possible to carry out an efficient simulation even in the infinite interaction range considered here~\cite{Schuch2008}. We find a similar behavior for the maximum entropy reached in the dynamics for the intermediate regime $0<\delta_\omega/\mu\lesssim 1$ as shown in panel(b). For non-zero values of the energy asymmetry however, the half-chain entanglement entropy $S_{N/2}(t)$ shows strong oscillations analogous to those found in similar spin models undergoing a dynamical phase transition~\cite{Roggero2021b}. We note that an increased dynamics of the entanglement entropy has been already associated in the past to the presence of a dynamical phase transition, but was usually associated with a increase of the time-derivative of $S_{N/2}$ at the critical times~\cite{Heyl2018,Haldar2020}.

The time $t_{ent}$ needed to reach the first maximum in the entanglement entropy is shown in panel (c) of Fig.~\ref{fig:1b_entropy_wt} and has three distinct scaling regimes: for $\delta_\omega<0$ is approximately independent from system size, at the critical point we recover the familiar $t_P\approx\sqrt{N}$ and for positive $\delta_\omega>0$ we find a logarithmic evolution $t_P\approx\log(N)$ which seems to approach a constant again for large value of $\delta_\omega$. In this regime the system seems to enter a second "gapped" phase, similar to the one for $\delta_\omega<0$, but simulations with larger systems might be required to confirm this.

We note at this point that the slow increase of entanglement entropy with system size (and with time) is precisely what allows us to perform a computationally tractable approximation to the time evolution using Matrix Product States~\cite{Vidal2003_mps,Schuch2008}. 

\section{Summary and Perspective}
\label{sec:conc}

Entanglement is a fundamental feature of interacting quantum many body systems~\cite{Vidal2003,Eisert2010} which, somewhat surprisingly, can be shown to play no important role in determining ground-state properties of systems with all-to-all interactions like for the neutrino forward-scattering Hamiltonian of Eq.~\eqref{eq:fs_hamilt} (see eg.~\cite{Brandao2016}). This is at the heart of the expectation that a mean-field treatment of the dynamics would also be appropriate. However, the role played by entanglement in out-of-equilibrium settings, like those relevant for flavor evolution in a supernova environment,
is not as well understood~\cite{Cervia2019,Rrapaj2020}.

In this work we have solved for the real time dynamics of a simpler system described by the Hamiltonian in Eq.~\eqref{eq:ham_massb} sharing many similarities to the model used to describe bipolar oscillations~\cite{Hannestad2006,Duan2010}. In order to overcome the exponential complexity of a direct solution of the problem we use a simple technique, based on a MPS representation~\cite{Vidal2003_mps}, whose computational complexity scales with the amount of entanglement generated by the dynamics. The results for this simple model suggest that the maximum bipartite entropy reached in collective neutrino oscillations scales only as $\log(N)$ with the logarithm of the system size. This favorable scaling allows for the efficient calculation of the dynamics for relatively long times and systems exceeding $100$ neutrinos. The computational efficiency could also be improved by using more complex tensor networks like MERA~\cite{Vidal2008} which naturally describe states with logarithmically increasing entropy.

As shown in an accompanying paper~\cite{Roggero2021b}, the behavior observed here is also found in similar (non-integrable) spin systems whose out-of-equilibrium dynamics crosses a quantum critical point, suggesting a strong link between dynamical phase transitions and fast collective oscillations. This link provides us directly with conditions, like Eq.~\eqref{eq:inst_cond}, to determine when instabilities can occur. These conditions provide a generalization of those commonly obtained with linear stability analysis to the long time, non-linear, regime.

The artificial setting considered here, with a strongly asymmetric initial state and no effect from the vacuum mixing angle, was chosen to highlight differences with a typical mean-field treatment of the problem where the dynamics is simply absent. We expect however the qualitative conclusions reached here, fast oscillations caused by a dynamical phase transition and a general slow growth of the entanglement entropy with system size, to be valid also in more general situations where multi-angle effects, and more complicated energy distributions are considered~\cite{Roggero2021b}. In this regime, MPS simulations could be used reliably to track flavor dynamics up to relatively large, and more realistic, neutrino systems. Important extensions of this work will be to include effects due to finite-mixing angles, the interaction with external matter, the generalization to the 3 flavor case and the more realistic time-dependent setting.

Based on the results presented here, we still cannot completely exclude the existence of neutrino configurations able to generate substantially more entanglement during the dynamics. Interestingly, the presence of these classically-hard configurations would be signalled clearly by the failure of the MPS method to scale efficiently to large system sizes. If found, these would be ideal models to explore using near-term quantum simulators like trapped-ion systems~\cite{Zhang2017} or future circuit based quantum computers~\cite{Hall2021}, which do not necessarily suffer from an exponentially increasing computational cost when large entanglement is present. In future studies, the effects discussed here will be integrated into more realistic simulations to better ascertain the role of entanglement in the dynamical evolution of dense neutrino environments of astrophysical importance.

\begin{acknowledgments}
I want to thank Joseph Carlson, Vincenzo Cirigliano, Huaiyu Duan, Joshua Martin, Amol Patwardhan, Ermal Rrapaj and Martin Savage for the many useful discussions about the subject of this work. This work was supported by the InQubator for Quantum Simulation under U.S. DOE grant No. DE-SC0020970 and by the Quantum Science Center (QSC), a National Quantum Information Science Research Center of the U.S. Department of Energy (DOE). 
\end{acknowledgments}
 
%

\appendix

\section{Analysis of time-step errors}
\label{app:tstep}

In this appendix we provide more information about the choice of time step in the MPS simulations used in the main text. We first recall the expression for the approximate evolution operator used in our simulations
\begin{equation}
U(\delta) = \prod_{i<j=1}^Ne^{-i\delta h_{ij}} \prod^{1}_{i<j=N} e^{-i\delta h_{ij}}\;,
\end{equation}
and the very conservative error bound (see e.g.~\cite{Childs2021} for tighter ones) presented in Eq.~\eqref{eq:time_approx} of the main text
\begin{equation}
\label{eq:short_time_error}
\left\|e^{-i\delta H} -U(\delta)\right\| < \delta^3 \|H\|^3\;,
\end{equation}
where $\|\cdot\|$ are operator norms. This approximation is then useful for short-time evolutions only and in order to perform accurate simulations for a long time $t$  it is necessary to break the interval $[0,t]$ in small steps of size $\delta\ll t$ as described in Eq.~\eqref{eq:full_prop} of the main text. Using the union bound, one can straightforwardly generalize Eq.~\eqref{eq:short_time_error} to arbitrary long times as follows
\begin{equation}
\label{eq:short_time_error}
\left\|e^{-it H} -U(\delta)^{\frac{t}{\delta}}\right\| < \frac{t}{\delta}\delta^3 \|H\|^3=t\delta^2\|H\|^3\;.
\end{equation}
This results implies that, for a fixed evolution time $t$, the error scales quadratically with the time step and that to guarantee a maximum error $\epsilon$ it is sufficient to choose
\begin{equation}
\delta \leq \sqrt{\frac{\epsilon}{t\|H\|^3}}\quad\text{or}\quad \frac{t}{\delta} \geq \frac{(t\|H\|)^{\frac{3}{2}}}{\sqrt{\epsilon}}\;,
\end{equation}
where $t/\delta$ is the total number of steps in the simulation.

\begin{figure}
 \centering
 \includegraphics[width=0.45\textwidth]{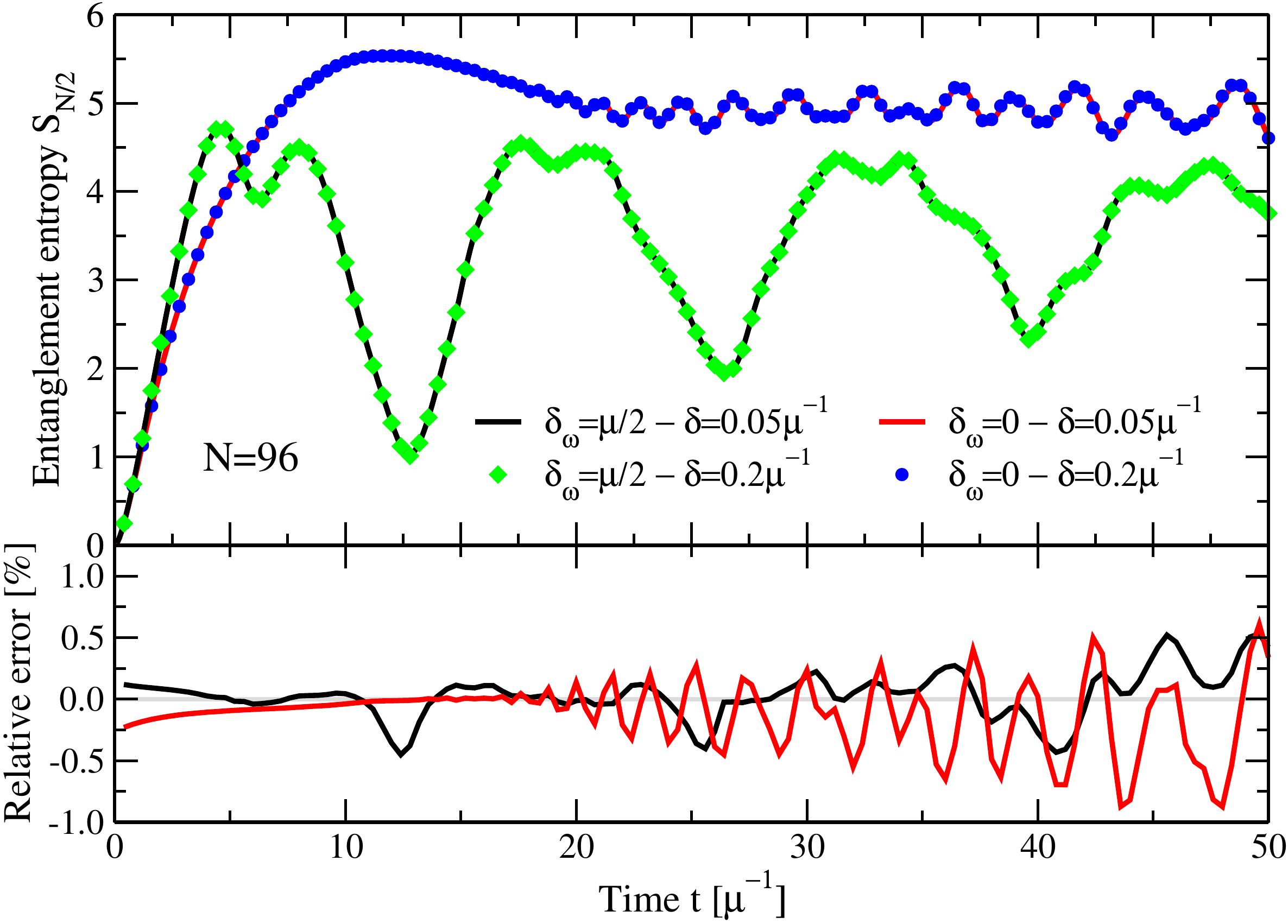}
 \caption{(Color online) Top panel shows the half-chain entanglement entropy in a system of $N=96$ neutrino amplitudes for two values of the one body asymmetry $\delta_\omega=0$ and $\delta_\omega=\mu/2$ and two different time steps $\delta=0.05\mu^{-1}$ (continuous lines) and $\delta=0.2\mu^{-1}$ (dots). The bottom panel shows the relative percent error caused by time step errors. }
\label{fig:ent_cmp}
\end{figure}

\begin{figure}
 \centering
 \includegraphics[width=0.45\textwidth]{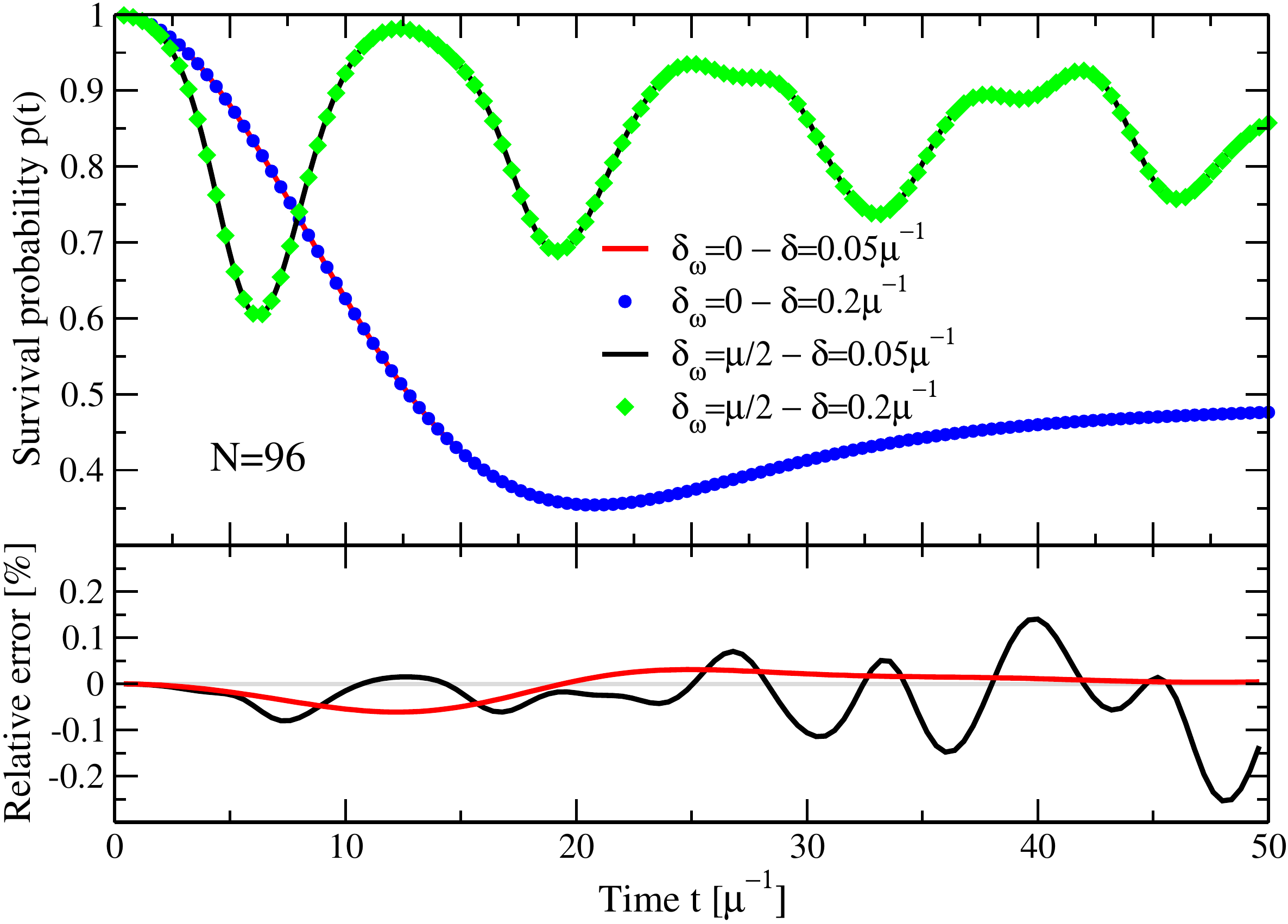}
 \caption{(Color online) Top panel shows the survival probability $p(t)$ in a system of $N=96$ neutrino amplitudes for two values of the one body asymmetry $\delta_\omega=0$ and $\delta_\omega=\mu/2$ and two different time steps $\delta=0.05\mu^{-1}$ (continuous lines) and $\delta=0.2\mu^{-1}$ (dots). The bottom panel shows the relative percent error caused by time step errors. }
\label{fig:pers_cmp}
\end{figure}

In order to estimate the contribution of this error to the observables displayed in the main text we performed additional simulations for the $N=96$ system at both $\delta_\omega=0$ and $\delta_\omega=\mu/2$ using a time step $\delta=0.2\mu^{-1}$ four times larger than the one used for the simulations described in the main text. We show the comparison between these two choices of time step for a system with $N=96$ neutrinos and two values of the energy asymmetry $\delta_\omega$ for the entanglement entropy in Fig.~\ref{fig:ent_cmp} and for the survival probability in Fig.~\ref{fig:pers_cmp}. The top panels show with continuous lines the results for $\delta=0.05\mu^{-1}$ (the choice used for the results in the main text) and the dots indicate the results with $\delta=0.2\mu^{-1}$ which are expected to have $\approx16$ times larger time-step errors. The observed deviations are extremely small for all times and, as shown in the bottom panels of both figures, increase slightly with the total evolution time but are always below $\approx 1\%$ for the entanglement entropy and $\approx0.2\%$ for the survival probability. Importantly, for times close to the maximum of the entropy the errors are well below the percent level indicating that the time step error is not a significant contribution to the uncertainty of the results presented in the main text.

\section{Direct comparison with mean-field}
\label{app:mf_cmp}
In this last section we show results obtained from a direct comparison between the MPS representation used in the work presented in the main text and a more standard mean-field solution of the evolution equations. This is done in order to better elucidate the role of the approximations used in the time-evolution algorithm described in the previous section and of possible finite-size effects which are absent in the mean-field.

The model considered in the main text is in the mass basis and in the mean-field approximation there is no flavor evolution. By rotating the initial state toward the flavor basis one can however produce collective oscillations also within the approximation. In this section, we will compare the flavor evolution in the mean-field approximation and the thermodynamical limit $N\to\infty$ with a MPS simulation at finite $N$ where at each time step we truncate the bond dimension to $r=1$ on all links. 

For this discussion we will use the same Hamiltonian in Eq.(3) of the main text, but rotated to the flavor basis
\begin{equation}
\label{eq:ham_flav}
H_f = \delta_\omega \vec{B}\cdot\left(\vec{J}^A-\vec{J}^B\right)+\frac{\mu}{N}\left(\vec{J}^A+\vec{J}^B\right)^2
\end{equation}
with $\vec{J}^A$,$\vec{J}^B$ the total spin operators in the two beams and $\vec{B}=(\sin(2\theta),0,-\cos(2\theta))$ is related to the mixing angle $\theta$. In the following, we will also keep the same initial condition $\ket{\Psi_0}$ as in the text, but now in the flavor basis. The equation of motion for the spin operators can then be expressed as follows
\begin{equation}
\begin{split} 
  \frac{d}{dt}\vec{J}_A &= \left[\delta_\omega\vec{B}+\frac{2\mu}{N}\left(\vec{J}_A+\vec{J}_B\right)\right]\times\vec{J}_A\\
 \frac{d}{dt}\vec{J}_B &= \left[-\delta_\omega\vec{B}+\frac{2\mu}{N}\left(\vec{J}_A+\vec{J}_B\right)\right]\times\vec{J}_B\;.
\end{split}
\end{equation}
One can eliminate the explicit dependence on the number of amplitudes $N$ by expressing these equations in terms of polarization vectors $\vec{P}_A = \frac{4}{N}\langle\vec{J}_A\rangle$ and $\vec{P}_B = \frac{4}{N}\langle\vec{J}_B\rangle$. In the mean field approximation we have then
\begin{equation}
\label{eq:mf_pvecs}
\begin{split}
  \frac{d}{dt}\vec{P}_A &= \left[\delta_\omega\vec{B}+\frac{\mu}{2}\left(\vec{P}_A+\vec{P}_B\right)\right]\times\vec{P}_A\\
 \frac{d}{dt}\vec{P}_B &= \left[-\delta_\omega\vec{B}+\frac{\mu}{2}\left(\vec{P}_A+\vec{P}_B\right)\right]\times\vec{P}_B\;.
\end{split}
\end{equation}
The solution of these coupled equations can be simplified with the change of variables proposed in Refs.~\cite{Duan2006b,Hannestad2006}. 

\begin{figure}[b]
 \centering
 \includegraphics[width=0.48\textwidth]{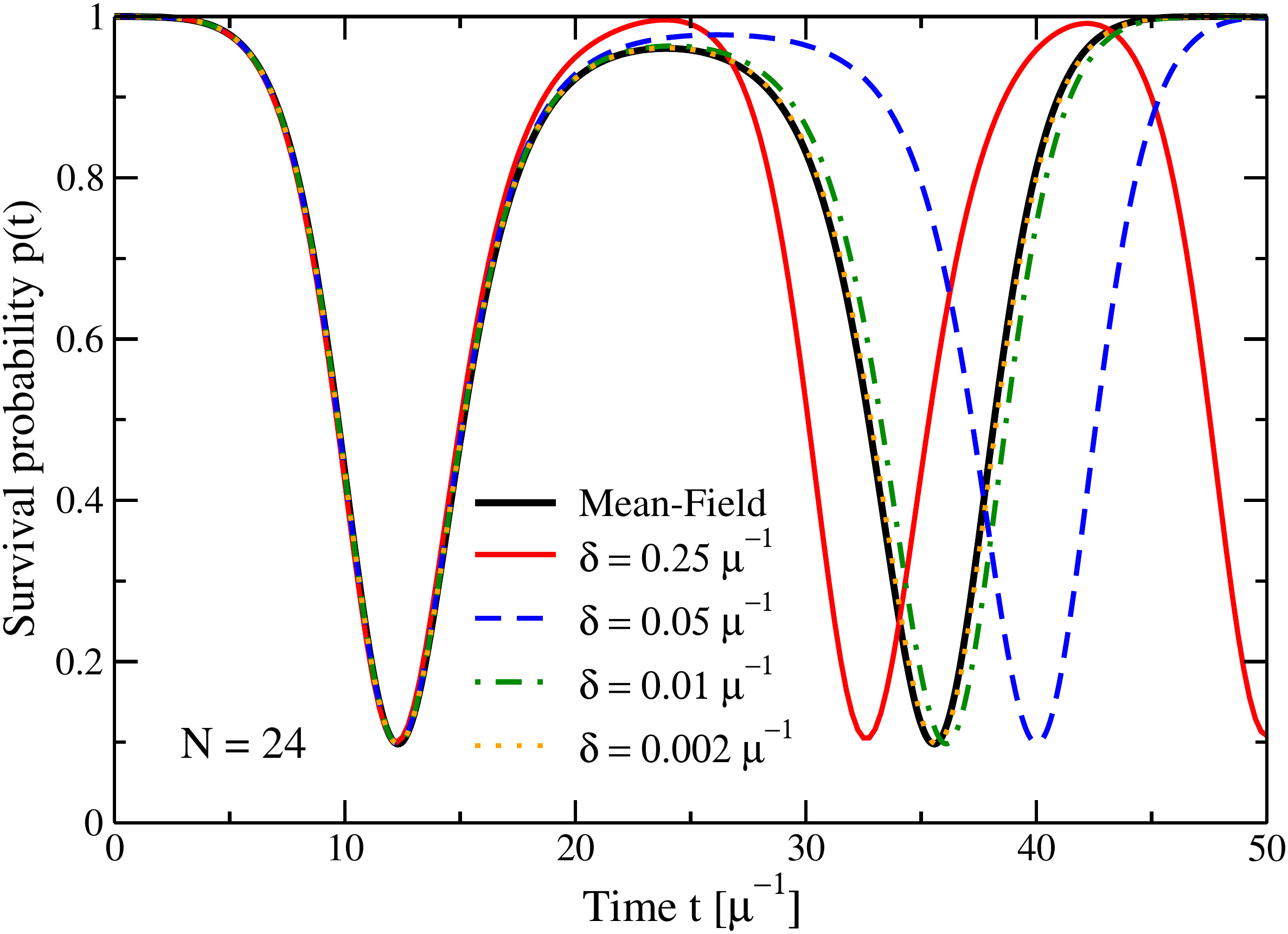}
 \caption{(Color online) Survival probability $P(t)$ as a function of time for a system with $N=24$, $\delta_\omega/\mu=0.1$ and bond dimension fixed at $r=1$ for different values of the time-step $\delta$. The solid black curve corresponds to the solution of the mean-field equations Eq.~\eqref{eq:mf_pvecs}.} 
\label{fig:mf_cmp}
\end{figure}

In Fig.~\ref{fig:mf_cmp} we compare the result for the time evolution of the survival probability
\begin{equation}
P(t) = \frac{1}{2}\left(1-\vec{P}_A\right)=\frac{1}{2}\left(1+\vec{P}_B\right)\;,
\end{equation}
as obtained by either solving the equations in Eq.~\eqref{eq:mf_pvecs} (black solid line) or using the MPS time-evolution scheme described above with a fixed bond dimension of $r=1$ (this figure is equivalent to the solid curve in Fig.~2(b) of Ref.~\cite{Duan2006b}, upon rescaling of the time variable). The different colored curves correspond to different choices for the time-step $\delta$ used to simulate time-evolution with Eq.~\eqref{eq:full_prop}. The results show a discrepancy with the mean-field results that grows as a function of time and depends on the chosen time-step. This effect is due to the approximate way the evolution is maintained at the mean field level: the local SVD-based truncation of a link tensor to $r=1$ does not guarantee to select the globally optimal tensor. This deficiency of the time evolution scheme described in Sec.~\ref{sec:mps_time_evol} is one of the main motivation behind alternatives like the Time Dependent Variational Principle(TDVP) approach~\cite{Haegeman2011} which, instead, approximates globally optimal tensors remaining directly on the manifold of matrix product states with fixed bond dimension. As the operators $G^{\sigma_1\sigma_2}_{\sigma_1'\sigma_2'}$ used in the updates from Eq.~\eqref{eq:updated_t} become closer to the identity, this truncation error becomes smaller. We can clearly see this effect in Fig.~\ref{fig:mf_cmp} as we lower the size of the time step. These results show the importance of checking convergence of results with respect to the choice of $\delta$ (as done here in Appendix~\ref{app:tstep}) and also the potential benefits of incorporating ideas like the TDVP scheme in future studies of neutrino oscillations with MPS.

\end{document}